\documentclass{article}

\usepackage[english]{babel}

\usepackage[letterpaper,top=2cm,bottom=2cm,left=3cm,right=3cm,marginparwidth=1.75cm]{geometry}

\usepackage{amsmath}

\usepackage[colorlinks=true, allcolors=blue]{hyperref}
\usepackage[pdftex]{graphicx}
\usepackage{float}

\title{ Payday loans - blessing or growth suppressor? - Machine Learning Analysis}
\author{
  Rohith Mahadevan\\
  \texttt{rohithmahadev30@gmail.com}
  \and
  Sam Richard\\
  \texttt{josephiterichard0411@gmail.com}
  \and
  Kishore Harshan Kumar\\
  \texttt{harshankumarhrk@gmail.com }
  \and
  Jeevitha Murugan\\
  \texttt{jeevithamurugan.2512@gmail.com}
  \and
  Santhosh Kannan\\
  \texttt{mail2santhoshkannan@gmail.com}
  \and
  Saaisri\\
  \texttt{saaisri2002@gmail.com}
  \and
  Tarun\\
  \texttt{tarun.s1997@gmail.com}
  \and
  Raja CSP Raman\\
  \texttt{raja@tactii.com}
}

\begin{document}
\maketitle

\begin{abstract}
The upsurge of real estate involves a variety of factors that have got influenced by many domains. Indeed, the unrecognized sector that would affect the economy for which regulatory proposals are being drafted to keep this in control is the payday loans. This research paper revolves around the impact of payday loans in the real estate market. The research paper draws a first-hand experience of obtaining the index for the concentration of real estate in an area of reference by virtue of payday loans in Toronto, Ontario in particular, which sets out an ideology to create, evaluate and demonstrate the scenario through research analysis. The purpose of this indexing via payday loans is the basic - debt: income ratio which states that when the income of the person bound to pay the interest of payday loans increases, his debt goes down marginally which hence infers that the person invests in fixed assets like real estate which hikes up its growth. 


\end{abstract}

\section{Introduction}

The most underrated factor that comes into play in the history of the real estate market in Toronto seems to be the payday loans whose APR (annual percentage rate) is much higher than other loans in Canada. In addition to the high-interest rates charged, the late payment fee and the NSF (non-sufficient fund - return check charge) fee are charged by the lender to the borrower in some cases of need. The eligibility for a person to borrow is at least 18 years old or the age of majority in a province or territory, have proof of an income and be a Canadian citizen or permanent resident. One must also need to have a valid bank account. Some lenders may ask for access to the borrower's bank account which often creates a hassle of the overdraft, which holds an ultimatum to the debt cycle, further the income of the person drains which results in the worst scenario of investment in fixed assets like real estate. 
In addition to it, the verge of increasing indebtedness causes a dramatic decline in the economic status of the citizens and for their upliftment and the social status, The Government of Canada offers 3 programs to help first-time homebuyers – the first-time homebuyer incentive, the first-time home Buyer’s amount tax credit, and Home Buyers’ Plan. However, this would help the citizens sufficiently. Still, a further increase in the interest turns the situation to be alarming and would decline the growth factor of the real estate market. 
Hence, the paper has undertaken the theme of indexing the upsurge of real estate via the reduction of the concentration of payday loans. Unknowingly, this domain of analysis stays farther from the views of the common man, who turns out to be the prey to the huge interest amount further expanding the debt cycle. Hitherto, there have been no signs of drastic growth of real estate in the area of payday loans concentration which turns out to be harsh both on the economy as well as the citizens.
Thereby, the upcoming parts of this paper explain in detail the analysis made using the datasets, unsupervised machine learning, data analytics, etc to come to a conclusion of the indexing satisfying the motto of the research paper.

\section{ Literature Survey:}

 \cite{san19}The highest level of financial literacy among Canadians has caused unusual growth in Canada over the years. The Canadian government implemented First-Time Home Buyer Incentive(FTHBI) to increase in Canadian ownership; as a result, people faced many financial, unemployment, and intersectional shifts in labour among various sectors. \cite{article}According to observations, the Canadian labour market has experienced rapid change in the last decade. Unemployment can result in intersectoral shifts in labour demand and the process of labour reallocation across sectors. The government has increased the payroll type of tax in recent years. Wages and salaries increased by approximately 10 percent, while supplement labour increased by 40 percent, increasing people's salaries by approximately 3 percent. This leads to inconsistency in consumption and borrowing, forcing these people to rely on micro-credit. \cite{NBERw28799} Furthermore, if borrowers are "naive" about their current focus, overly optimistic about their future financial situation, or do not anticipate their high likelihood of repeat borrowing for any other reason, they may underestimate the costs of repaying a loan. In this case, restricting credit access may benefit.\cite{gi21} Credit constraints and financial exclusion are used to achieve this. Income, assets or wealth, age, household size, education, debt, gender, and region of residence are used as predictors of financial exclusion. \cite{article2}This information is obtained using credit reports which contain a wealth of information about consumers' financial habits. Hundreds of consumer-related variables are evidently collected in these reports from a variety of sources. \cite{gajo} Although most customers were unaware that being denied a loan had an impact on their credit history, learning this increased their interest in providing an indication of loan eligibility on the site. Loans were compared based on both total cost and Annual Percentage Rate(APR). While APR was not widely understood, it was perceived as simple to compare and served as a proxy for price when the total repayment amount was not specified. \cite{al22}  Because of geographic variation in access to loans caused by regulation, the effects of payday loans on consumers are very mixed. There are more prisoners in the area as a result of the crisis. Regressions were used to examine the lender's credit history. This provides adequate lending and lender information.\cite{br09} Network data adds value by increasing the predictive accuracy of these models. Predictors are intended to inform decisions that affect the very outcomes that are the target of prediction, such as when the predictor is intended to estimate the risk of an adverse outcome so that a decision-maker can intervene to prevent that outcome. A predictor trained to be fair in one population may be unfair in another, but performativity can induce unfairness even when the populations in which the predictor is trained and deployed are the same, i.e. there is no covariate shift. \cite{st94} Thus, two major conclusions can be reached. To begin, future research should concentrate on addressing each stage of the payday loan process, beginning with understanding the needs of payday loan users. The findings indicate that borrowers' decisions are close to optimal in light of their liquidity needs; however, the initial liquidity needs that drive people to seek payday loans may be the result of suboptimal consumption and savings decisions.

\section{DATA ARCHITECTURE:}
 \begin{figure}[H]
    \centering
    \includegraphics[width=\linewidth]{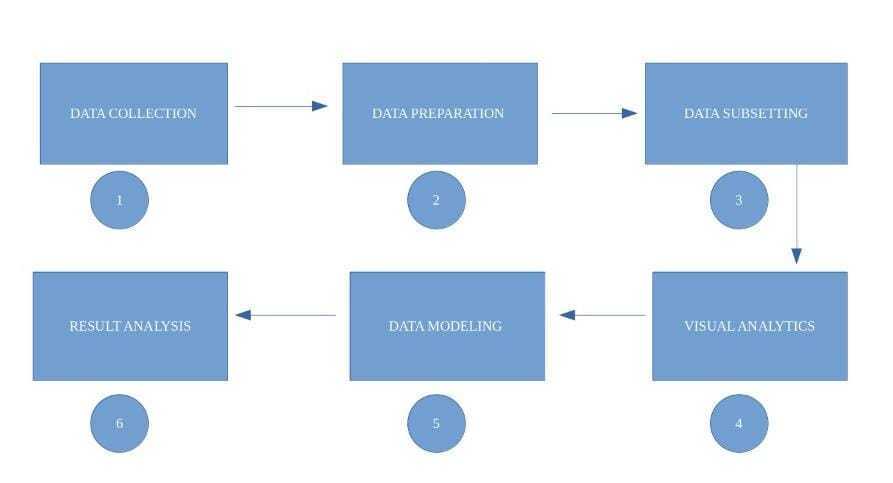}
    \caption{Flow chart on data architecture}
    \label{fig:my_label2}
 \end{figure}
\section{Methodology }
\subsection{ Data collection:}
Initially, the data is gathered from numerous sources on the internet to determine the number of payday loan lenders in each location based on facts such as 2 bedrooms, 3 bedrooms, and 4 bedrooms in 2010 and 2020 among various regions of Canada. For a high level of accuracy, the data was cross-verified numerous times. such that this data is more adequate for further data preparation
\subsection{Data preparation:}
Before removing null values from structured data, the data was properly cleansed. The clustered data was then structured into a single data frame for easier identification, and an average analysis was returned for 2010 and 2020. For a better comprehension of data, numerous visualizations are used. The data was further divided into several number of subsets and visualized accordingly. Further, it is proceeded with various algorithms to reach better accuracy. 
\subsection{Inference from analytics:}
Analytics is the process that consists of data cleaning, inspecting, modelling, and transforming the data to obtain useful insights from which conclusions can be drawn. Analytics is invaluable to the process of decision-making and optimization. Modelling and visualizing are two major aspects of analytics that help us to better understand the data and utilize resources effectively.

Initially, the data is divided into sub-sections for analysis purposes. After, cleaning the data and preparing it, statistical analysis is done. Initial observation of how many payday loan lenders are present in each city in the Ontario province is seen. 
\begin{figure}[h!]
    \centering
    \includegraphics[scale=0.5]{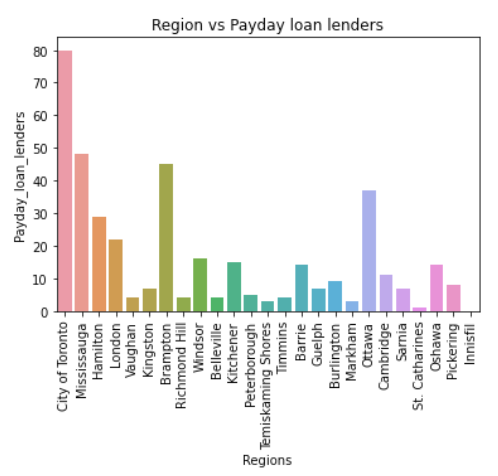}
    \caption{Graph indicates payday loan lenders with reference to various regions of Canada}
  
    \label{figure 2}
\end{figure}
\subsection{Statistical inference:}

The value in the real estate price is calculated by using the percentage increase formula. The highest percentage increase of 861 percent is observed in the 4-bedroom apartment. The real estate price increased from 2010 to 2020 drastically in specific areas. An average increase of 128 percent is observed in the overall data.deep analysis, it is inferred that the mean increase was very less in 2 bedroom apartment.
\begin{figure}[h!]
    \centering
    \includegraphics[scale=0.5]{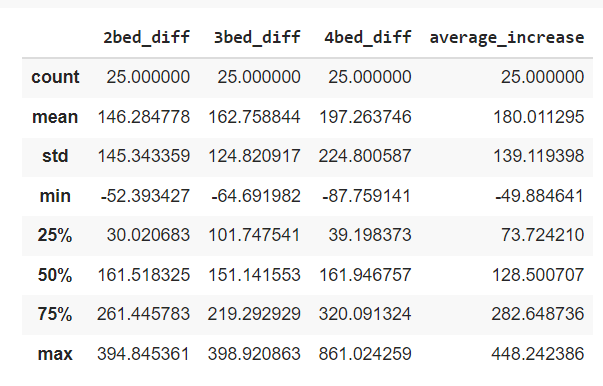}
    \caption{Analysis on  dataset}
    \label{fig:my_label3}
\end{figure}
Even though the dataset is cleaned, some data from 2010 are approximated values. The visualization is done with the present data that are taken from multiple resources. The data is taken from up-to-date websites that are found online. A result comparison of real estate prices in 2010 and 2020 is done. The chart shows that there is an overall increase but in some areas, there is a high increase in prices. In some areas, the prices are increased moderately with an increase in time. For further drill-down analysis, the data is cleaned in such a way as to model the efficiency based on the available data. 
\begin{figure}[h!]
    \centering
    \includegraphics[scale=0.5]{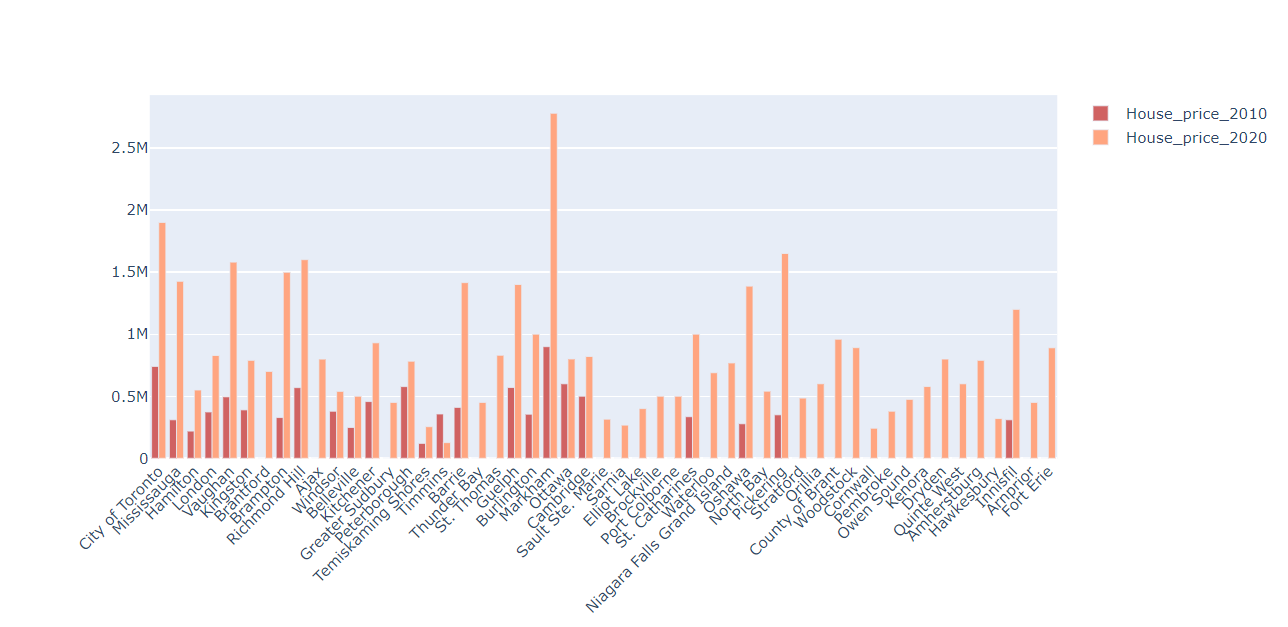}
    \caption{House prices in 2010 and 2020}
    \label{fig:my_label4}
\end{figure}
The median of the price increase is inclined towards the first quartile region. This means that there is a huge increase observed over the period 
\begin{figure}[h!]
    \centering
    \includegraphics[scale=0.5]{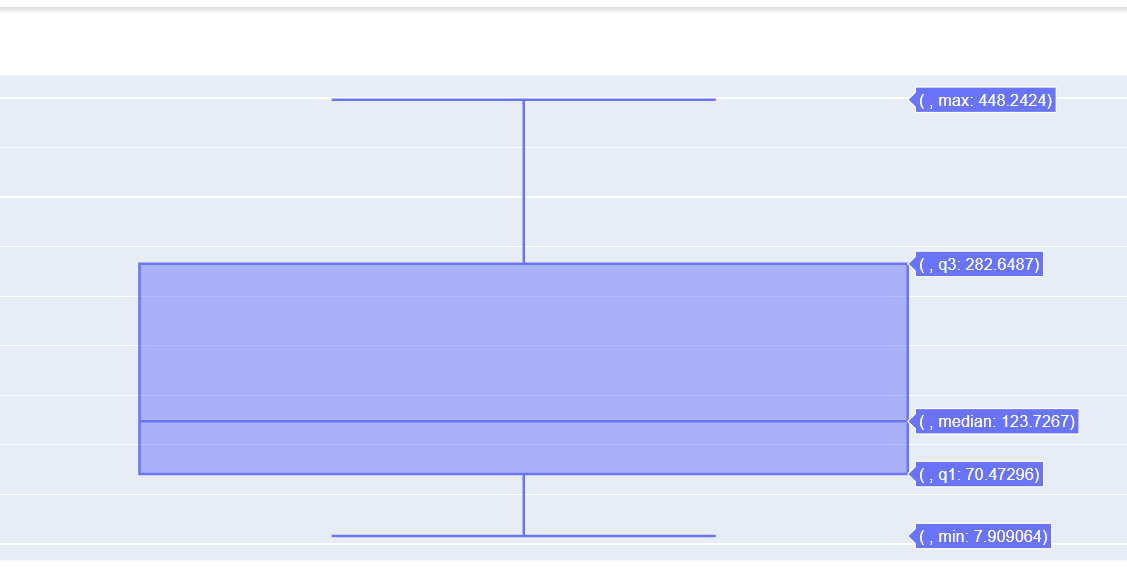}
    \caption{Box plot analysis on Median of increase in price }
    \label{fig:my_label5}
\end{figure}
\subsubsection{Regression analysis}
The data from the dataset is preprocessed and a simple plot is made. Different combination of average increase vs payday loan lenders, 2 bedroom price percentage increase vs payday loan lenders, 3 bedroom price percentage increase vs payday loan lenders, 4 bedroom price percentage increase vs payday loan lenders is made. Considering all the combinations, 4 bedroom houses show that the percentage increase was in receding trend. The trend was normal and predicted a downfall
\begin{figure}[h!]
    \centering
    \includegraphics{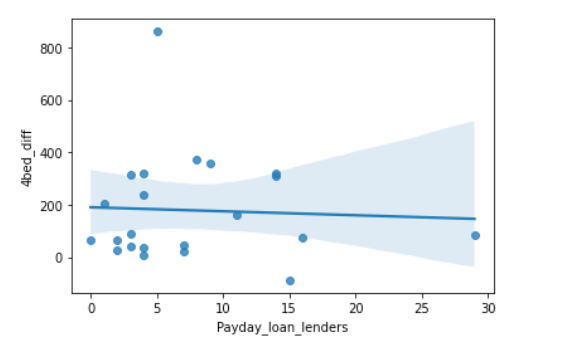}
    \caption{Analysis of payday loan lenders vs average increase in pricing }
    \label{fig:my_label5}
\end{figure}

The chart analysis interprets that as the payday loan lender's percentage increases the price variation in real estate price decreases. The downtrend graph proves the statement. For further analysis, k means clustering is done to find whether there are any cluster of groups formed with respect to payday loan lender's and the average percentage increases in price. 
\section{Data Modeling}
\subsection{Percent increase formula:}
\begin{center}
 percent increase =\( \frac{final-initial}{|initial|}{\cdot100}\)
\end{center}
\subsection{K-Means Clustering:}
K-Means Clustering is a simple unsupervised learning algorithm that is used to divide the dataset into K number of distinct, non-overlapping clusters. It divides the data into clusters based on the centroid of the cluster. A cluster's centroid is the arithmetic mean of the values in the cluster. A data point is allocated to a cluster if the squared distance between the data point and the cluster’s centroid is a minimum. The value of K is the hyperparameter in this algorithm that is generally found using the elbow and the silhouette methods.

The K means clustering algorithm is implemented in the dataset. The elbow method is used to find the number of clusters required to train the dataset . In the Elbow method, we are varying the number of clusters ( K ) from 1 – 10. For each value of K, we are calculating WCSS ( Within-Cluster Sum of Square ). WCSS is the sum of the squared distance between each point and the centroid in a cluster. When we plot the WCSS with the K value, the plot looks like an Elbow. As the number of clusters increases, the WCSS value will start to decrease. WCSS value is largest when K = 1. When the graph is analyzed it is evident that the graph will rapidly change at a point and thus creating an elbow shape. From this point, the graph starts to move almost parallel to the X-axis. The K value corresponding to this point is the optimal K value or an optimal number of clusters. 
\subsection{Implementation of K-Means clustering: }
The algorithm is implemented using the python sklearn library. The elbow method determines the number of clusters that are actually needed for the analysis. The n clusters that are required are 3 and the data is trained with the algorithm. After training, the data points are split into three different groups. As discussed, a cluster is formed where the payday loan lenders are less and the average increase in percentage is increasing.  
\begin{figure}[h!]
    \centering
    \includegraphics[scale=0.3]{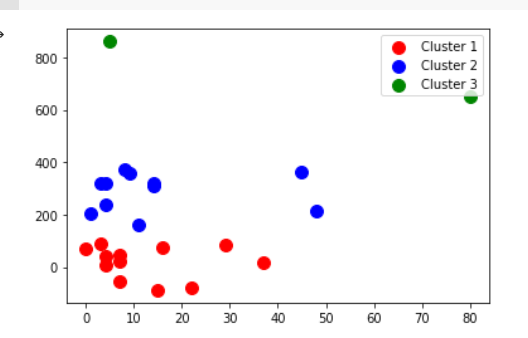}
    \caption{Analysis plot on three different subsets}
    \label{fig:my_label}
\end{figure}

This supports our previous regression analysis. The average trend tends to be in a state of descending. 
\section{Result analysis:}
The interpretation of the graphical representation starts with the normalization of the data before
visualizing it which plays a pivotal role in the data analysis. The dataset  involves the pricing of 2,3,4 Bedroom houses in the Ontario province and is estimated. A percentage over the difference of the range is obtained and plotted in the graph. The graph is now visualized using a regression plot .

Overall, this analysis shows that there is a gradual decrease in the growth factor of real estate by virtue of payday loans. However, the graph is plotted against the payday loan lenders and the average increase of the houses which leads to a marginally decreasing slope which is welcoming to support our primary thesis. 

Further, anomaly has to be checked wherein there is one single anomaly which is inferred that is the shift of the interquartile range which states that all the outcomes of the cluster blobs come under the position and have one anomaly data in 4Br. To figure this out and to give out a precise outcome we take into consideration the k means clustering from the unsupervised machine learning which involves the division of the entire graphical representation into small parts of subset ; say over a range of 0-30. The quartile 3 of the graph has more outcome markings as compared to the quartile 1 because of that the region of q3 is more as compared to q1. More the cluster in the graph less is the payday loans the median shifts in the q1 region. 

The elbow chart developed from the k means cluster analysis shows the kirks/ bents in the decreasing slope that is interpreted, thereby the random state is declared,if k increases, the payday loan lenders will decrease, and each cluster will now have fewer constituent instances and the instances occur closer to the respective centroids. However, the increase in payday loans will decline the growth of real estate where k is taken into consideration. 
\section{Limitations of the study:}
The hypothesis that exists in the real world is that payday loan lenders indirectly affect the real estate growth of a particular area. The research paper correlates all the data that are available on the internet and tries to possibly prove the hypothesis. The research study is concentred only on the Ontario province. The data analytics and modelling are done for the dataset which is concentrated towards the specific region thus resulting in very minimal insights. The problem with the dataset is all the real estate prices from 2010 are not available as there were fewer data collected at that time. A crucial takeaway from this result is that we can't fully foresee the impact of real estate growth on payday loans because the land or surroundings may not meet the expectations of consumers, causing real estate growth to spike.
\section{Conclusion :}
This research paper mainly focuses on Ontario province and the cities in it the above study concludes that amongst many factors influencing the growth rate of real estate; payday loan lenders too  play a significant role. The research analysis concludes that the increase in payday loans would marginally decrease the real estate growth in particular; which in general may involve a variety of factors too. 
However, the study stated above involves likely the views of the small group of  researchers while the participants of the paper do not claim or may not be representative of larger segments of populations; which may challenge traditional findings, implications, generalizability, and measures of validity. 
\section{Future work:}
The future work of this research paper is to concentrate on all the provinces in Canada and to find more insights into the impact of payday loans on people. Thus, this will strengthen the study and will help us to find more about the indirect impact of the payday loan concept and how it affects the real estate growth in the country. The study will also concentrate on the factors which have an indirect impact on the economic growth of the country and possibly try to increase the overall economic status of the common people living across the country. 
\section{Acknowledgements:}
We'd like to express our gratitude to our Featurepreneur team for collecting data to improve this model's accuracy. Thank you for mentoring the paper Snekha Suresh.

\bibliographystyle{unsrt}
\bibliography{sample}

\end{document}